\documentclass{aa}
\usepackage{graphicx}
\usepackage{natbib}
\usepackage{amssymb}
\begin{document}

\title{Observational clues for a role of circumstellar accretion in PMS X-ray 
activity}
\author{E. Flaccomio \and G. Micela \and  S. Sciortino}
\offprints{E. Flaccomio,
\email{ettoref@oapa.astropa.unipa.it}}

\authorrunning{Flaccomio et al.}
\titlerunning{Circumstellar accretion and activity in the PMS}

\institute{INAF - Osservatorio Astronomico di Palermo Giuseppe S. Vaiana,
Palazzo dei Normanni, I-90134 Palermo, Italy\\
\email{ettoref@oapa.astropa.unipa.it, giusi@oapa.astropa.unipa.it, sciorti@oapa.astropa.unipa.it}}

\date{Received date / accepted date}

\abstract{ 
We revisit the published analyses of ROSAT X-ray observations of the star
forming regions NGC~2264 and Chamaeleon I ($\sim 3$ and $\sim 5$ Myr old
respectively) in the light of newly published optical data. At odds with
previous results on Chamaeleon I members, we find that low mass stars in both regions have near-saturated emission levels. Similarly to what previously
found in the Orion Nebula Cluster, Weak Line T-Tauri Stars in NGC~2264 and in
the Chamaeleon I cloud have higher X-ray activity levels respect to Classical T
Tauri Stars, arguing in favor of a role of the disk and/or accretion in
determining X-ray emission.  

}

\maketitle

\section{Introduction \label{sect:intro}}

The influence of accretion disks surrounding young PMS stars on their observed
X-ray activity levels is presently debated. The topic has been investigated
many times through imaging X-ray observations of star forming region but
contradictory results are reported. Mentioning just a few recent examples,
\citet{fei02} analyze {\em Chandra} ACIS-I data finding no indication that the
presence of an accretion disk modifies activity levels of Orion Nebula Cluster
(ONC) stars. The same negative result, although somewhat controversial, is
reported for IC~348 members by \citet{pre01,pre02}, also using {\em Chandra}
ACIS-I observations; by \citet{law96} for the Chamaeleon I cloud using ROSAT
PSPC data; by \citet{fla00} for NGC~2264 using the ROSAT HRI; by \citet{gro00}
for $\rho$ Ophiuchi again with the ROSAT HRI; by \citet{get02} for NGC~1333
(ACIS-I).

On the other hand, Classical T-Tauri Stars (CTTS) belonging to the
Taurus-Aurigae association are found to be sub-luminous in the X-ray band
respect to Weak Lined T-Tauri Stars (WTTS) by both \citet{neu95} and
\citet{ste01}. \citet{fla02b}, using  {\em Chandra} HRC-I data, report a
similar result, with high statistical confidence and at odds with
\citet{fei02}, for the rich ONC population. Other indications of a difference
between CTTS and WTTS have been found in X-ray band variability characteristics
and spectra. Namely, \citet{ste00} in Taurus-Aurigae, \citet{fla00} in NGC~2264
and Flaccomio et al. (in preparation) in the ONC, all find that CTTS are more
variable than WTTS. Some studies have also indicated that CTTS may have
different X-ray spectral characteristics respect to WTTS: \citet{tsu02} find
that the mean $kT$ for CTTS is about 3 keV, compared to $\sim 1.2$ for WTTS.
Such a large $kT$ difference may in part be due to a selection effect: in the
X-ray selected sample of \citet{tsu02} class II sources (CTTS) are
significantly more absorbed respect to class III-MS sources (WTTS) and it is
therefore possible that only the hardest CTTS have been observed. Other
contrasting indications have been also presented: \citet{kas02}, using high
resolution X-ray spectra of the 10Myr old CTTS TW Hydrae, derive a differential
emission measure distribution peaking at $\sim 0.3$ keV and propose that the
emission mechanism is related to matter accretion. No systematic difference in
$kT$ between CTTS and WTTS is observed by \citet{pre02} in IC~348 members.

Are these contradictory results due to real differences between different star
forming regions or to the different approaches used in analyzing and
interpreting data? We will touch upon four important points that can affect the
result: 1) accounting for the mass/$L_{\rm bol}$ dependence of PMS activity; 2)
choosing a relevant accretion/disk indicator; 3) avoiding selection effects in
the reference stellar sample; 4) converting observed X-ray photon detection
rates to X-ray luminosities. 

\paragraph{Mass/$L_{\rm bol}$ dependence.} It is possible that the failure to
detect a difference in activity levels between stars with different
circumstellar/accretion properties  
is due to the fact that the activity levels are also influenced by other stellar characteristics, and the various contributions have not been disentangled.
In particular a
dependence of mean $L_{\rm X}$ on stellar mass (or bolometric luminosity) has been
widely found for PMS stellar groups. Most of the studies mentioned above
compare the X-ray Luminosity Function (XLF) of CTTS with that of WTTS, both
XLFs being computed from stellar samples comprising a wide range of masses.
Such a procedure tends to hide possible differences because: 1) the presence
and/or magnitude of the effect sought might depend on stellar mass; 2) if a
different mass-$L_{\rm X}$ relation holds for these two classes, stars having the same
$L_{\rm X}$, but different mass, will contribute to both XLFs. A more sensible
approach, in order to eliminate this source of confusion (see e.g.
\citealt{fla02b}), is to compare XLFs of stars in restricted mass ranges; this
however requires, for meaningful statistical comparisons, sufficiently large
samples of well characterized members. 
To the same effect, considering that the ratio $L_{\rm X}/L_{\rm bol}$ is, 
for low mass PMS stars ($\lesssim 3M_\odot$), on average much less
dependent from mass than $L_{\rm X}$ (e.g. \citealt{fla02b}), it is also sensible 
to compare the distributions of this latter parameter for the two classes. 
As an added advantage, $L_{\rm X}/L_{\rm bol}$ is arguably less
sensitive to interstellar extinction corrections (although the newly
introduced variable, $L_{\rm bol}$, also carries non-negligible uncertainties).
Both of these approaches were successfully followed by \citet{fla02b} to
establish the difference in activity levels between accreting and non-accreting
ONC members.

\paragraph{Disk/accretion indicator.} There is no widespread consensus on which
indicator of presence of disk or of accretion is to be used to search for
effects on activity levels. Some studies have correlated X-ray data with {\em
accretion} indicators, such as the $H_{\rm \alpha}$ or $Ca~II$ line emission. Others
have instead employed {\em circumstellar disk} indicators such as near IR excesses
(in K or L band). The relation between presence of disks and matter accretion
phenomena is not yet fully understood. A statistical correlation between
accretion and disk indicators is generally observed but it seems clear that not
all IR detected disks are associated with accretion and it is also possible
that the presence of accretion is not always related to a near-IR detectable
disk (e.g. because of a large inner disk hole that suppress the K band excess
or because of the disk view-angle). It is presently unclear whether X-ray
emission levels are influenced by one (or both) of the two phenomena. It seems
therefore reasonable to explicitly distinguish between the two. As a clarifying
example, \citet{pre02} find statistically significant evidence that accreting
stars in IC~348 have lower $L_{\rm X}$ respect to non accreting ones, but no evidence
of a difference between stars with and without $K-L$ color excess (a disk
indicator). They consider these two results contradictory, maintain that {\em the
infrared excess gives a more realistic picture of the circumstellar properties
of the T Tauri stars than the $H_{\rm \alpha}$ emission}\footnote{The reason \citet{pre02} do
not trust $H_{\rm \alpha}$ as an indicator are its time variability and the fact that
part of the $H_{\rm \alpha}$ flux can be of chromospheric origin. However variability
should if anything tend to lower the significance of the difference between
CTTS and WTTS and the chromospheric origin of part of the $H_{\rm \alpha}$ flux
should at most produce an effect opposite to that observed: the sample of
strong $H_{\rm \alpha}$ stars, interpreted as accreting stars, will be indeed
contaminated by chromospherically active stars, which are also likely to have
high coronal activity and thus $L_{\rm X}$.} and attribute the detected difference in
X-ray luminosity functions to selection biases. An alternative point of view
would be that {\em accretion} and not the presence of an {\em IR-detectable
disk} influences X-ray activity and the two are not simply related. We note
that in other cases in which a difference in activity levels is reported the
distinction was performed on the basis of accretion indicators: $H_{\rm \alpha}$ in
Taurus (e.g. \citealt{ste01}) and $Ca~II$ in the ONC \citep{fla02b}.

\paragraph{Parent sample selection.} Ideally, a complete and not-contaminated
sample of members of a given SFR should be used to investigate the matter.
However, this is in practice hardly possible. Understanding selection biases is
therefore crucial: in the case of IC~348 discussed above, \citet{pre02} suspect
that strong, easily detectable $H_{\rm \alpha}$ emission may favor the inclusion in
the reference stellar sample of optically (and X-ray) faint accreting stars and
therefore artificially depress the mean $L_{\rm X}$ of CTTS. This, i.e. selecting
members on the basis of their circumstellar/accretion properties, or in any
other way that favors the selection of faint CTTS over that of WTTS, is indeed
the main risk to be avoided, or accounted for, in such a study. Taking the
approach of dividing the whole sample in narrow mass ranges (see above), this
problem, usually worse for low mass, low $L_{\rm X}$ stars, is reduced for the higher
mass ranges. Other member selection methods, are not likely to result in
spurious results: selection through sensitivity limited X-ray observations, for
example, will sample to the same minimum $L_{\rm X}$ both CTTS and WTTS (assuming
similar X-ray spectra and absorptions), so that if the two underlying (i.e.
complete) XLFs do not differ, the detection fraction of both classes will be
the same and the two distributions of observed $L_{\rm X}$ will not differ either.
Inclusion in the reference sample of contaminating non-members, usually low
$L_{\rm X}$ stars, will depress the mean activity levels inferred for non
accreting/disk-surrounded stars, therefore going in the direction of producing
the opposite result respect to the observed one.

\begin{figure}[t]
\centering
\resizebox{\hsize}{!}{\includegraphics{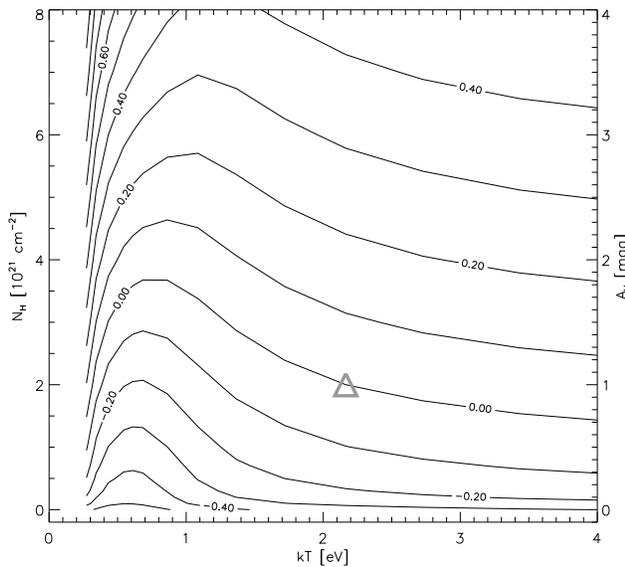}}
\caption{Contour plot of the logarithm of the conversion factor ($CF$) from
{\em ROSAT} HRI count-rates to unabsorbed flux, in the 0.1-4.0keV spectral
band,  for coronal thermal sources (Raymond-Smith emission model) as a function
of source temperature, $kT$, and absorption, $N_{\rm H}$. Lines refer to loci in
which $Log(CF)$ is constant and equal to the value reported in the
corresponding label plus -10.07, i.e. the $Log(CF)$ corresponding to {\em
typical} parameter values: $kT=2.16$,  $N_{\rm H}=2\cdot 10^{21}$ (marked in the
Figure by a triangle). Line spacing is 0.1 dex in $Log(CF)$. The ordinate axis
on the right gives the {\em standard} conversion between $N_{\rm H}$ and optical
absorption, $A_{\rm V}$: $N_{\rm H}=2\cdot 10^{21}A_{\rm V}$.} 
\label{fig:app_inst_cf}
\end{figure}

\paragraph{X-ray count-rate to $L_{\rm X}$ conversion.} X-ray telescopes detect
photons in wide energy ranges. The conversion between detected count-rate and
X-ray luminosity depends therefore on the incoming spectra. Given the low
statistic of most X-ray sources with present day instruments and/or the lack of
spectral resolution of some X-ray detectors, it is necessary to assume a source
spectrum.\footnote{Even in the case of detectors with intrinsic spectral
resolution, the low source counts practically prevent in many cases the
determination from the data of a reliable spectral model.} For coronal sources
this usually reduces to assuming the $kT$ of a thermal emission spectrum and
the hydrogen column density, $N_{\rm H}$, of the absorbing material between the
source and the observer. Differences in the way these two parameters are
estimated can lead to significantly different conversion factors and therefore
affect the result of our search for a difference between CTTS and WTTS. On one
hand, systematic  differences in the X-ray spectra ($kT$ and $N_{\rm H}$) of CTTS and
WTTS might result, if not properly accounted for, in spurious results regarding
the different X-ray luminosity of the two classes.\footnote{Differences in the
intrinsic spectra would however be interesting "per se" for the understanding
of the physical mechanism that determines X-ray emission.} On the other hand,
significant random errors in the conversion could easily wash out an existing 
correlation between $L_{\rm X}$ and accretion/disk indicators. In the works of
\citet{fei02} and \citet{get02}, for example, the evidence presented against a
difference of WTTS and CTTS is obtained using detector plane (non
absorption-corrected) X-ray fluxes. Likewise \citet{law96} and \citet{fla00},
also obtaining a negative result, assume a single count-rate to flux conversion
factor and therefore neglect any difference in absorption between sources.
\citet{fla02a,fla02b} on the other hand, although assume a single $kT$ for all
sources, correct for individual absorption values ($N_{\rm H} \propto A_{\rm V}$), finding
a positive result. \citet{ste01} also correct for absorption (through a low
energy hardness ratio) and find a positive result. Is it possible that a
peculiarity in the $N_{\rm H}$ vs. $A_{\rm V}$ relation or in the intrinsic spectra of CTTS
and WTTS results in artificially lowering the luminosities derived for CTTS
respect to those derived for WTTS? Regarding the relation between X-ray and
optical extinction, recent studies correlating $N_{\rm H}$, derived from X-ray medium
resolution spectra, and optically derived $A_{\rm V}$ confirm the relation between
the two and do not evidence any such difference \citep[cf.
][]{ima01,fla02a,fei02,koh02}. It is however possible, although presently not still
fully established, that CTTS have harder X-ray spectra respect to WTTS
\citep{tsu02}. We may wonder how a different $kT$ would affect the luminosities
we derive. Figure \ref{fig:app_inst_cf} shows the value of the conversion
factors for the {\em ROSAT} HRI, as a function of $kT$ and $N_{\rm H}$. Qualitatively
similar plots are obtained for the {\em ROSAT} PSPC and for the {\em Chandra}
HRC-I, the two other instruments used for the observations discussed later in
this paper. We observe that, for a given source $N_{\rm H}$, the difference in $kT$,
if eventually confirmed, will indeed go in the direction of decreasing the
inferred $L_{\rm X}$ of lower $kT$ sources respect to high $kT$ ones, thus
potentially accounting for part of the observed differences between CTTS and
WTTS.  However it is also clear that for typical $A_{\rm V} \sim 0.5-4.0$ the mistake
committed in not accounting for individual source temperatures could be at most
of the order of $\sim 0.1-0.2$ dex, smaller than the difference between CTTS
and WTTS found by \citet{fla02b} in the ONC. We stress however that even if
this effect were to be confirmed, therefore reducing the actual difference in
luminosities respect to that inferred assuming a single $kT$, the difference
between the X-ray emission of CTTS and WTTS would be confirmed, and the
spectral differences would provide additional clues for the understanding of
its physical origin. 

\bigskip

In this paper, keeping the above four points in mind, we further discuss and
extend the evidence for a role of accretion and/or disk in determining the
observed X-ray activity level of ONC members, as already reported by
\citet{fla02b}. In the light of newly available optical/IR data we then
critically reanalyze the results obtained by \citet{fla00} and \citet{law96},
both of which concluded that stars surrounded by disks, in NGC~2264 and Cha I
respectively,  have the same activity levels as those that do not have a disk.
Here we derive the opposite result. 

The structure of this paper is as follows: in  Sect. \ref{sect:ONC} we discuss the
new observational evidence for a difference in activity levels between CTTS and
WTTS belonging to the ONC. In  Sect. \ref{sect:2264} and Sect. \ref{sect:ChaI} we
then discuss the cases of NGC~2264 and the Chamaeleon I cloud. Finally in Sect.
\ref{sect:summary} we briefly summarize our results.

\section{The ONC: an update \label{sect:ONC}}

Our analysis method throughout this paper is based on the work of
\citet{fla02a,fla02b} on the ONC. This cluster is arguably the best target
available for our study because we have access to a rich and well characterized
sample of members spanning a wide range of masses. We refer the reader to
\citet{fla02a,fla02b} for a full description of the X-ray and optical data used
here. The extinction limited sample ($A_{\rm V} < 3.0$) discussed in this latter work
has little field contamination and is {\em complete} almost down to the lowest
stellar masses. Using the $Ca~II$ line ($\lambda = 8542\AA$) as an indicator  for
circumstellar accretion, \citet{fla02b} obtained with high statistical
significance the result that low mass stars ($M \lesssim 3 M_\odot$) with this
line in strong emission ($EW < -1$) have systematically lower $L_{\rm X}$ and
$L_{\rm X}/L_{\rm bol}$ values respect to stars with the line in absorption ($EW > 1$). 

\begin{figure*}[t]
\centering
\includegraphics[width=\textwidth]{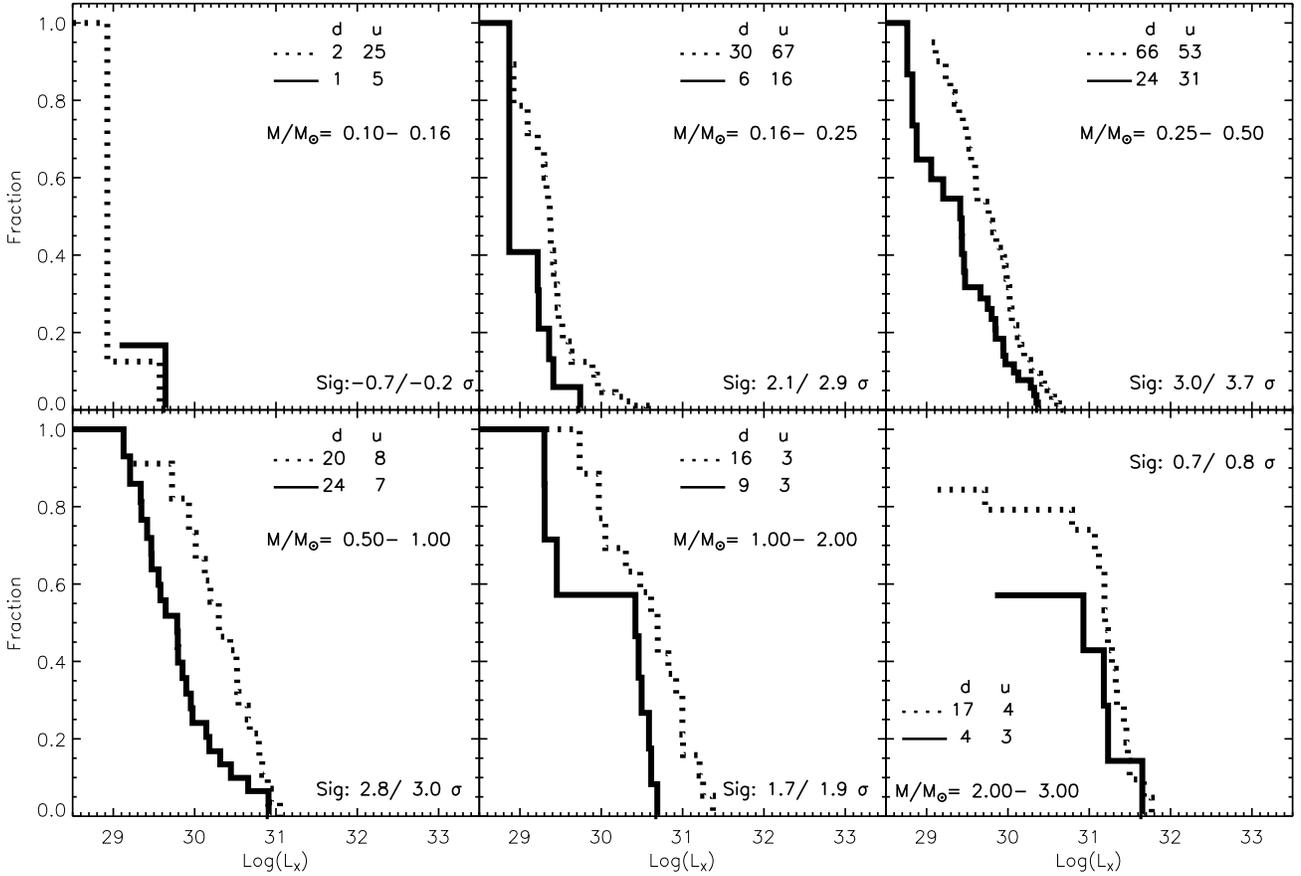}
\caption{X-ray luminosity functions for stars with high- and low-NIR excess (solid
and dashed lines, respectively) in the Orion Nebula Cluster. Panels refer to different mass ranges, as indicated
by legends. Also reported are the numbers of detected (d) and undetected stars (u)
used for XLFs of high- and low-accretion subsamples and the $\sigma$-equivalent significance range for the difference between the two distributions (see text). \label{fig:XLF_CW_ORI}}
\end{figure*}

\begin{figure*}[t]
\centering
\includegraphics[width=\textwidth]{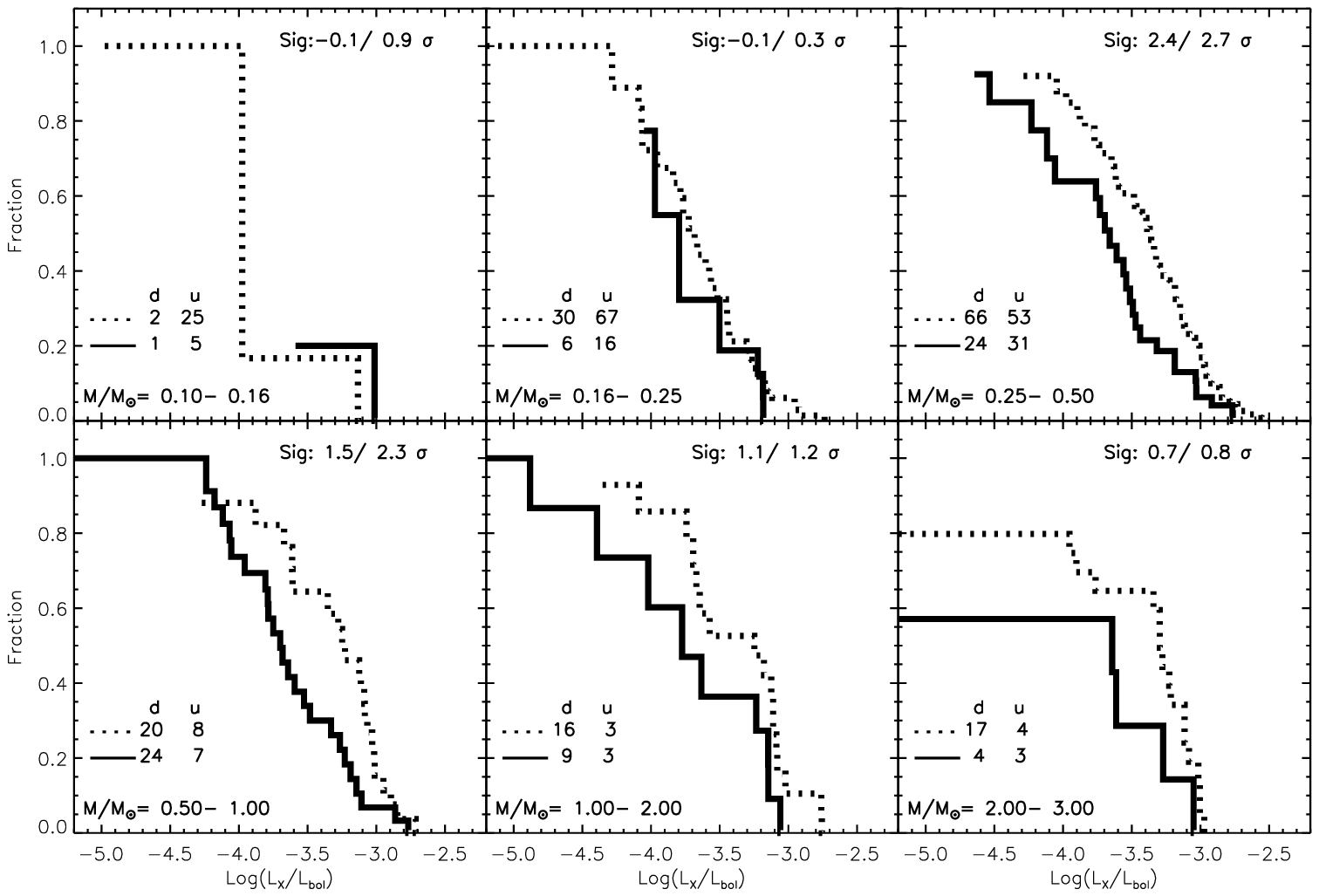}
\caption{Same as Figure \ref{fig:XLF_CW_ORI} for $L_{\rm X}/L_{\rm bol}$. \label{fig:LXLBOL_CW_ORI}}
\end{figure*}

Here we state that an analogous result is obtained, albeit with smaller
significance ($2.5-3 \sigma$), comparing the $L_{\rm X}$ and $L_{\rm X}/L_{\rm bol}$
distributions of stars with large and small near IR excess ($\Delta(I-K) > 0.8$
and $\Delta(I-K) < 0.4$ respectively). The X-ray and optical/IR data are
presented in \citet{fla02a} and the $\Delta(I-K)$ values are taken from
\citet{hil98}. Figure \ref{fig:XLF_CW_ORI} and \ref{fig:LXLBOL_CW_ORI} shows
the maximum likelihood $L_{\rm X}$ and $L_{\rm X}/L_{\rm bol}$ distributions for these two
classes of stars in 6 different mass bins. The range of confidence with which
we can exclude that the two distributions are randomly extracted from the same
parent population, according to the tests in the {\sc asurv} package
\citep{fei85}, is given inside each panel.

We also note that very similar results are obtained, both with the $Ca~II$ line
and with $\Delta(I-K)$ as a discriminant, if only X-ray detected members are
considered in the distribution functions. As reminded in the introduction,
these latter results exclude that the difference in the distributions is due to
the preferential selection of faint CTTS (which is anyway not expected given
that the sample is {\em not} selected from either accretion or disk
indicators).

\section{NGC~2264 \label{sect:2264}}

Stellar activity in NGC~2264 has been studied most recently by \citet{fla00}
through 6 different {\em ROSAT} HRI observations covering, in two different
pointings, a large fraction of the star forming cloud. 
One hundred sixty nine 
distinct sources
were detected, $\sim 95\%$ of which are estimated to be associated with members
of the association. One of the main problem at the time of this work was the
lack of good optical characterization of the members, so that, for example,
lacking individual measurements, extinction toward sources had to be assumed
uniform and placement of counterparts in the HR diagram was performed solely
from optical photometric data, an error-prone procedure for PMS stars. Moreover
the distinction between CTTS and WTTS was not well established, as indications
on the NIR excesses were not available and $H_{\rm \alpha}$ measurements were in most
cases qualitative and non-uniform. Since that work new improved optical data
have been recently published by \citet{reb02}. We therefore updated the
previous analysis according to the general principles stated in the
introduction. 

\subsection{The reference sample}

We adopted optical data from tables 1 and 3 of \citet{reb02}. Out of the full
list of 687 photometrically selected candidate members (i.e. the {\em in cloud,
in locus} population defined by \citealt{reb02}) we selected the 202 stars for
which reliable spectral types and extinction ($A_{\rm V}$) estimates (through $R$ and
$I$ photometry $+$ spectral types) were available. This latter is a subset of
the full spectroscopic sample studied by \citet{reb02}, initially selected
primarily from a list of I-band variable stars, with the addition of previously
known candidate members based on their X-ray or $H_{\rm \alpha}$ emission or on their
proper motion. Our reference sample is therefore the intersection (logical {\em
and}) of the photometrically selected member sample and of the spectroscopic
sample. While the  former is arguably free from selection biases in favor of
faint CTTS, the degree of representativeness of the latter in this respect is
less clear: $H_{\rm \alpha}$ is however only a secondary selection criterion and
I-band variability (periodic in $> 50\%$ of the stars), although maybe more
frequent in CTTS, does not obviously favor the inclusion of optically faint
stars. Moreover the disk (CTTS) fractions \citet{reb02} derive for the
spectroscopic and the photometric sample (cf. their Table~6) are remarkably
similar, suggesting that the former is not strongly biased toward CTTS.

Contamination of our reference sample from field stars may on the other hand be
non negligible: according to preliminary proper motion data \citet{reb02}
report that $\sim 50\%$ of their photometric candidates are actually
non-members. The spectroscopic sample, selected on the basis of PMS stellar
characteristics, is expected to be less contaminated, although an estimate
based on proper motion data is not provided. We recall (cf. Sect.
\ref{sect:intro}) that field star contamination is expected to artificially
lower the activity levels of WTTS.

We place stars in our reference sample in the HR diagram. Effective
temperatures and bolometric corrections are estimated from spectral types and
\citet{ken95} conversions.\footnote{The effective temperature for spectral type
A6 was modified from 8350 K to 8050. The former value seems to be at odds with
the spectral type-$T_{\rm eff}$ relation; we suspect a typographical mistake.} We
then evaluate bolometric luminosities from I band magnitudes. Out of the 202
spectrally characterized candidate members, 193 fall within the \citet{sie00}
evolutionary model grid and have therefore been assigned a mass and an age. 

The reference sample used for our following analysis comprises the 178 stars,
out of these 193 candidate members characterized in terms of mass and age, that
fall in the field of view of the X-ray observations described by
\citet{fla00}. 

\subsection{X-ray data}

We matched the photometric catalog of \citet{reb02}, out of which our reference
member list is drawn, with the list of 169 X-ray sources published by
\citet{fla00}. The identifications were carried out as described in 
\citet{fla00}, i.e. assuming as identification radii the the off-axis dependent
X-ray source position error summed in quadrature to 1$^{\prime\prime}$, i.e. a
conservative estimate of the optical position error.  Before performing the
final identification we first registered the coordinate systems of the optical
and X-ray lists by comparing the positions of 125 uniquely identified pairs
($RA_{\rm X}-RA_{\rm opt}=-1.7^{\prime\prime}$, $Dec_{\rm X}-Dec_{\rm opt}=1.9^{\prime\prime}$). 
Sixty seven
stars in our reference sample were identified with an X-ray source, 56 of which
uniquely, while the the remaining 11 fell in the identification circle of an
X-ray source along with other objects in the photometric catalog.  To each of
the 67 candidate members with X-ray counterparts we then assigned a Maximum
Likelihood (ML) X-ray count rate: these are values computed by \citet{fla00} in
order to define a mean source brightness among 6 different observations. Due to
source variability, five of these mean values are actually upper limits. The
count rates of the 11 ambiguous identifications were also treated as upper
limits and upper limits, computed as in \citet{fla00}, were also assigned to
111 X-ray undetected candidate members lying within the FOV of the HRI
observations.

\begin{figure}[t]
\centering
\resizebox{\hsize}{!}{\includegraphics{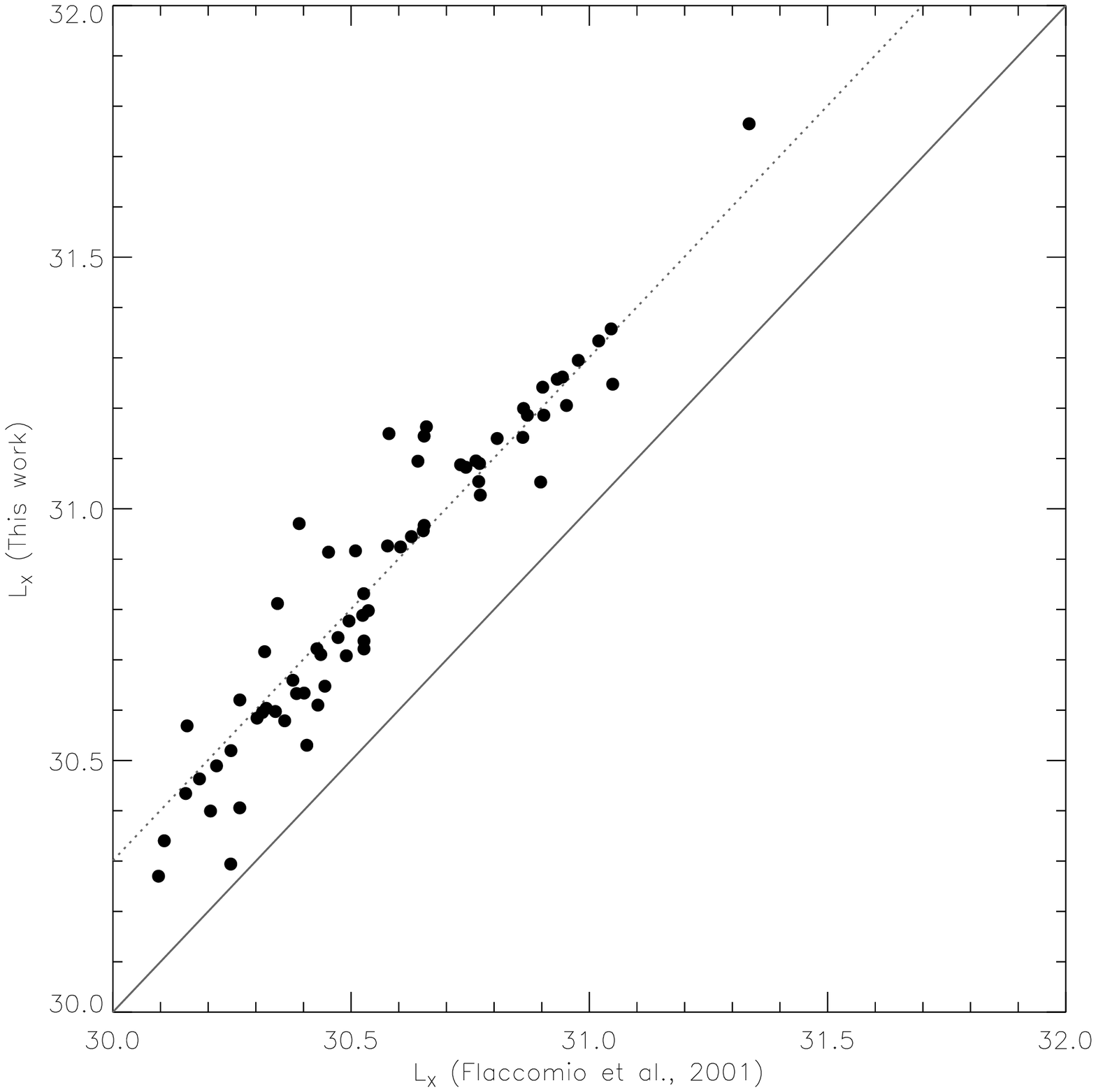}}
\caption{Comparison of X-ray luminosities computed by \citet{fla00} for NGC~2264 stars (see text) and those recomputed from the same data in this work. No distinction is made here between detections and upper limits. The solid line indicates the locus of equal values; the dotted lines indicate the relation $L_{\rm X}$(This work) $=2 \cdot L_{\rm X}$\citep{fla00}. The scatter of points about the mean relation is due to the adoption of individual extinction corrections.}
\label{fig:conp_2264_lxlx}
\end{figure}

Finally we converted count-rates, measured and upper limits, to X-ray
luminosities in the 0.1-4.0 keV band\footnote{Conversion factors to un-absorbed
flux are computed using PIMMS (Portable, Interactive, Multi-Mission Simulator),
version 3.0, available on-line at: http://asc.harvard.edu/toolkit/pimms.jsp}.
We assumed a thermal emission spectrum with $kT=2.16$ keV, close to recent
estimates for PMS stars \citep[e.g.][]{fla02b,ima01,get02}. The hydrogen column
density was assumed proportional to the optical extinction measured
individually for each star: $N_{\rm H}=2\cdot 10^{21} A_{\rm V}$. The distance was assumed
to be 760pc like in \citet{fla00}. Figure \ref{fig:conp_2264_lxlx} compares the
$L_{\rm X}$ derived by \citet{fla00} to those derived here from the same count-rates.
Our new estimates are $\sim 0.3$ dex higher respect to the old ones, with the
main differences due to the assumed $kT$ (2.16 vs. 0.75 keV) and the increased
absorption (mean $A_{\rm V} \sim 0.45$ vs. a constant $A_{\rm V}=0.19$) and a small
difference, $\lesssim 0.1$ dex, due to the different spectral band in which
$L_{\rm X}$ is computed.

\subsection{Activity vs. circumstellar environment}

We will now try to establish whether the activity levels of our sample of
NGC~2264 candidate members depend on circumstellar properties, evidenced either 
by accretion or by IR disk indicators. We will use two indicators provided by
\citet{reb02}: $H_{\rm \alpha}$ equivalent width, $EW_{\rm H\alpha}$, and excess in the
$H-K$ near IR color, $\Delta(H-K)$. In the former case we will consider as {\em
accreting} those stars with $EW_{\rm H\alpha} > 5$ and {\em non-accreting} those
with $EW_{\rm H\alpha} < 5$\footnote{We choose 5 instead of the more customary 10
as $EW$ threshold because we so obtain a better distinction between activity
indicators of the two classes. As noted in the introduction lowering the
threshold could, if anything, result in a reduction of the inferred difference
in activity levels; the most active WTTS might indeed have strong chromospheric
$H_{\rm \alpha}$ emission and be mistaken for CTTS.}. In the latter case we take as
threshold $\Delta(H-K)=0.15$ with stars showing larger excesses considered as
surrounded by circumstellar disks.

\begin{figure}[t]
\centering
\resizebox{\hsize}{!}{\includegraphics{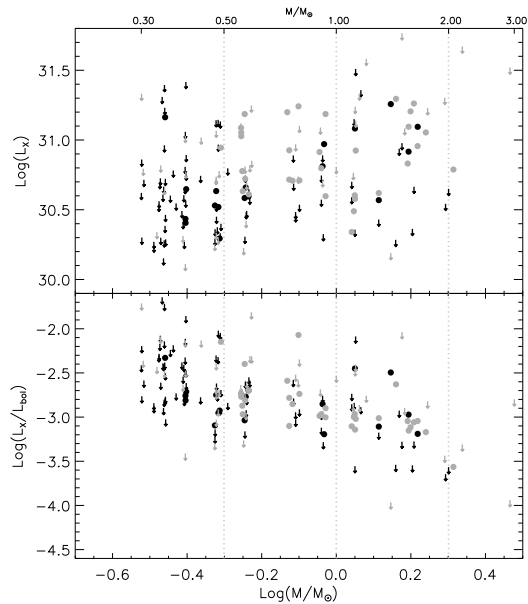}}
\caption{$L_{\rm X}$ and $L_{\rm X}/L_{\rm bol}$ vs. mass for NGC~2264 stars with high and low $H_{\rm \alpha}$ equivalent width (black and
and gray symbols, respectively). Filled circles represent detections, down-pointing arrows upper-limits.  \label{fig:act_mass_2264} }
\end{figure}

\begin{figure*}[!h!]
\centering
\includegraphics[width=\textwidth]{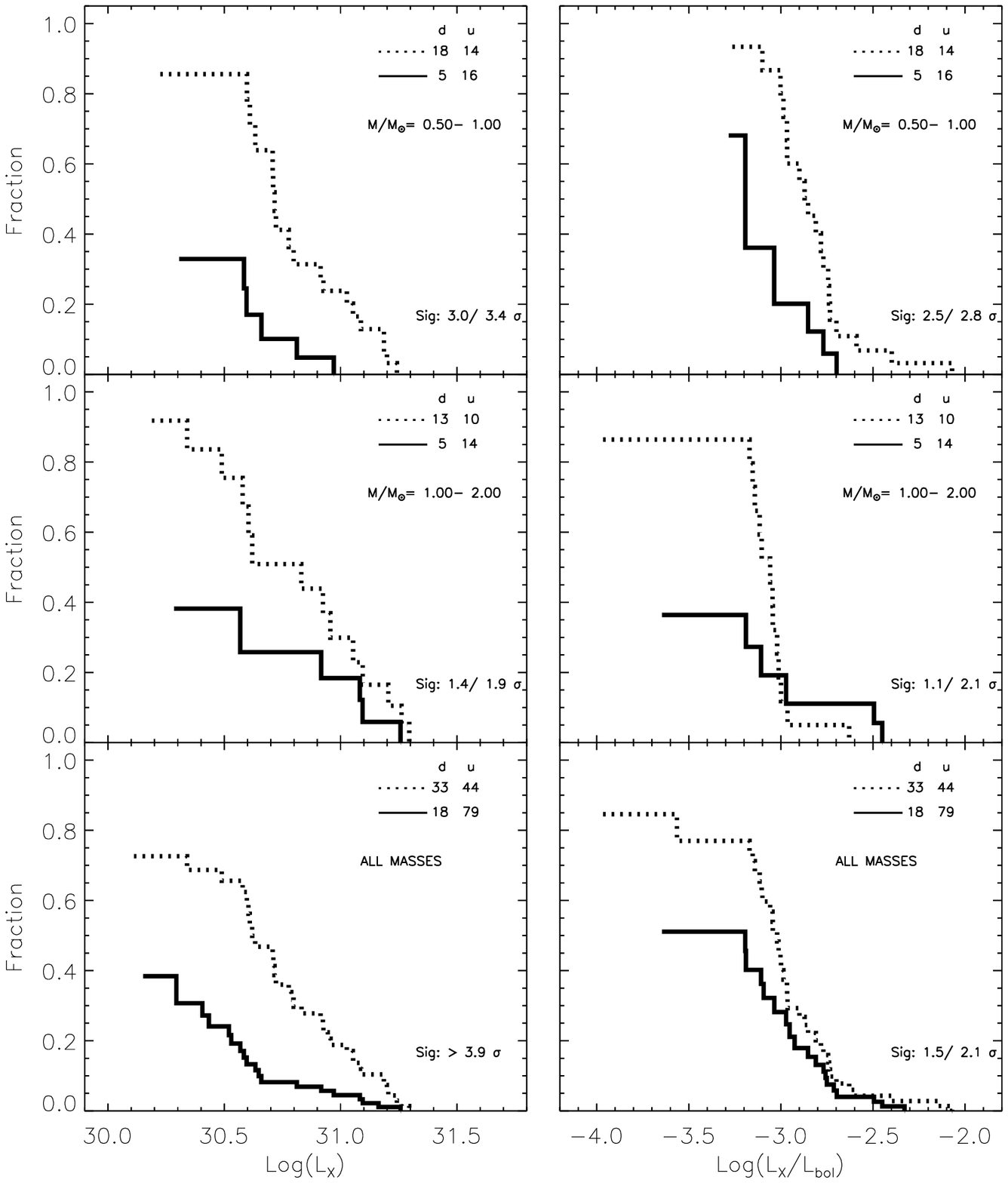}
\caption{Distributions of $L_{\rm X}$ and $L_{\rm X}/L_{\rm bol}$ (left and right columns) for NGC~2264 stars with high and low $H_{\rm \alpha}$ equivalent width (solid
and dashed lines, respectively). Each row refers to a different mass range as indicated. Legends inside panels as in Fig. \ref{fig:XLF_CW_ORI}. \label{fig:XLF_CW_2264}}
\end{figure*}

Figure \ref{fig:act_mass_2264} shows the scatter plots between $L_{\rm X}$ and
stellar mass and $L_{\rm X}/L_{\rm bol}$ and mass, for the accreting and non-accreting
stars. Similarly to what seen in other SFRs, $L_{\rm X}$ is seen to correlate with
stellar mass, although the relation appears somewhat fuzzier in this case
respect to, e.g., the ONC or the Chamaeleon I cases (see below). A systematic
difference between the two classes is not readily apparent. However, given the
large number of upper limits a more quantitative analysis is needed. Figure
\ref{fig:XLF_CW_2264} shows XLFs and $L_{\rm X}/L_{\rm bol}$ distributions for the two
$H_{\rm \alpha}$ separated stellar classes, in two different mass ranges, marked in
Fig. \ref{fig:act_mass_2264} by vertical lines, and for the whole sample. In
all cases the distributions of accreting stars appear to lie below those for
non-accreting ones. Statistical test (in the {\sc asurv} package) confirm such
differences with varying degree of confidence (results of the tests are given
in the figure, along with the number of detection and upper limits that enter
in the distributions; the same information is repeated in the first part of
Table~\ref{tab:CWTTS_sign}, where the results of other tests described below
are also reported). Particularly significant is the difference in $L_{\rm X}$ for the
whole sample ($> 3.9\sigma$). Such a difference might however result from a
larger fraction of low mass, and low $L_{\rm X}$, accreting stars (see Fig. 
\ref{fig:act_mass_2264}). In the two narrower mass ranges, the result is
however retrieved, with greater than $3\sigma$ confidence in the
$0.5-1.0M_\odot$ mass range. The difference is also observed in $L_{\rm X}/L_{\rm bol}$. 
Note that, as pointed out above, our reference sample may be significantly
contaminated by field stars,  and this would tend to depress the non-accreting
stars distributions, thus lowering the significance of the result. If we repeat
the same analysis including only stars confirmed as members by their IR excess
($\Delta(I-K) > 0.3$ or $\Delta(H-K) > 0.15$), $H_{\rm \alpha}$ emission ($EW > 5$)
and X-ray detection, we indeed find even more significant differences.
Particularly interesting are the differences in $L_{\rm X}/L_{\rm bol}$, because they are
less likely to be influenced by selection effects, a concern for this latter
restricted sample. Table~\ref{tab:CWTTS_sign} reports the results of these
tests. 

Table~\ref{tab:CWTTS_sign} also reports the results of the comparisons between
the stars with and without near IR excess, for the same mass ranges and the two
stellar samples described above. The same results is retrieved: stars showing
a $\Delta(H-K)$ excess, indicating the presence of a disk, have lower activity
levels respect to the complementary sample.

\begin{table*}
\caption{Significance of difference between CTTS and WTTS in NGC~2264}
\begin{tabular}{lrrrr|cc}
\hline
Mass $[M_\odot]$  &$N_{\rm d}(C)$& $N_{\rm u}(C)$&$N_{\rm d}(W)$& $N_{\rm u}(W)$& $Sign.(L_{\rm X})$ & $Sign.(L_{\rm X}/L_{\rm bol})$ \vspace{0.2cm} \\ 
\hline \hline
\multicolumn{7}{l}{$EW(H_{\rm \alpha})$ - {\em Whole sample}}  \\ 
\hline
0.5-1.0   	&  5 	 &	16 &	18  &	 14  &  {\bf 3.0/3.4}  & 2.5/2.8   \\
1.0-2.0   	&  5  	 &	14 &	13  &	 10  &       1.4/1.9   & 1.1/2.1   \\
All       	& 18  	 &	79 &	33  &	 44  &  {\bf $>$3.9}   & 1.5/2.1   \\
\hline
\multicolumn{7}{l}{$EW(H_{\rm \alpha})$ - {\em Confirmed members}} \\
\hline
0.5-1.0   	&  5  	 &	16 &	18  &	  2  &  {\bf $>$3.9} &{\bf 3.2/$>$3.9}  \\
1.0-2.0   	&  5  	 &	14 &	13  &	  1  &   2.1/2.7     &1.9/{\bf 3.1}   \\
All       	& 18  	 &	79 &	33  &	  7  &  {\bf $>$3.9} &{\bf 3.3/$>$3.9}  \\
\hline
\multicolumn{7}{l}{$\Delta(H-K)$ - {\em Whole sample}} \\
\hline
0.5-1.0   	&  1  	 &	 9 &	18  &	 20   &  1.3/1.8   &{2.2/\bf 3.2}  \\
1.0-2.0   	&  1  	 &	10 &	14  &	 11   &  1.9/2.2   & 2.1/2.5   \\
All       	&  4  	 &	30 &	40  &	 71   &  2.2/2.4   &{2.6/\bf 3.2}  \\
\hline
\multicolumn{7}{l}{$\Delta(H-K)$ - {\em Confirmed members}}  \\
\hline
0.5-1.0   	&  1  	 &	 9 &	18  &	  9   &  1.8/2.3   &2.6/{\bf 3.5}  \\
1.0-2.0   	&  1  	 &	10 &	14  &	  2   &  2.4/2.8   &2.9/{\bf 3.4}  \\
All       	&  4  	 &	30 &	40  &	 44   &  2.8/{\bf 3.0} &{\bf 3.4/$>$3.9}  \\
\end{tabular}
\label{tab:CWTTS_sign}
\begin{list}{}{}
\item[Note --] Description of columns: (1) Mass range; (2) Number of detected CTTS; (3) Number of CTTS with upper limits; (4) Number of detected WTTS; (5) Number of WTTS with upper limits; (6) Range of significance (expressed in $\sigma$ equivalent) for the difference between the $L_{\rm X}$ distributions of CTTS and WTTS according to the tests in ASURV. Results greater than $3\sigma$ are in boldface; (7) like (6), but for $L_{\rm X}/L_{\rm bol}$
\end{list}
\end{table*}

\section{Chamaeleon I \label{sect:ChaI}}

Our main source of data regarding the Chamaeleon I association is the work by
\citet{law96}: we use their member list (117 stars in their Table~B1) and their
estimates of $L_{\rm bol}$ and $T_{\rm eff}$ (available for 78 stars, Table~6).  We
adopt here a distance to the Chamaeleon I cloud of 160pc \citep{whi97,wic98},
20pc larger than the distance assumed by \citet{law96} in estimating bolometric
luminosities. We therefore increased the $L_{\rm bol}$ values accordingly. Masses
of 71 candidate members were derived from placement in the HR diagram and
interpolation of \citet{sie00} evolutionary tracks.  

The selection of candidate members in \citet{law96} is performed mainly on the
basis of either their $H_{\rm \alpha}$ or X-ray emission. The danger of
preferentially selecting faint stars (both optically and in X-rays) with strong
$H_{\rm \alpha}$ emission is therefore present. However the Chamaeleon I association
is close enough that a large fraction of intermediate mass members is probably
detected in the {\em ROSAT} PSPC X-ray observations. Also,  as anticipated in
the introduction, other than $L_{\rm X}$ we will also investigate the dependence of
$L_{\rm X}/L_{\rm bol}$ on circumstellar characteristics and, as a further test, we will
also consider a fully X-ray selected sample. 

X-ray data were taken from \citet{law96}: they quote X-ray luminosities (or
upper limits) for members of the region, computed from ROSAT PSPC count rates
in the 0.4-2.5keV spectral band \citep{fei93}, using a constant count-rate to
$L_{\rm X}$ (in the same band) conversion factor: 1 PSPC count $\rm
ks^{-1}$=$3\times10^{28} \rm ergs \ s^{-1}$ . \citet{fei93} find that this
conversion factor corresponds to assuming a plasma temperature $kT \sim 1$keV
and an absorption by a hydrogen column, $N_{\rm H}$, corresponding to $A_{\rm V} \sim 1$.

\begin{figure}[t]
\centering
\resizebox{\hsize}{!}{\includegraphics{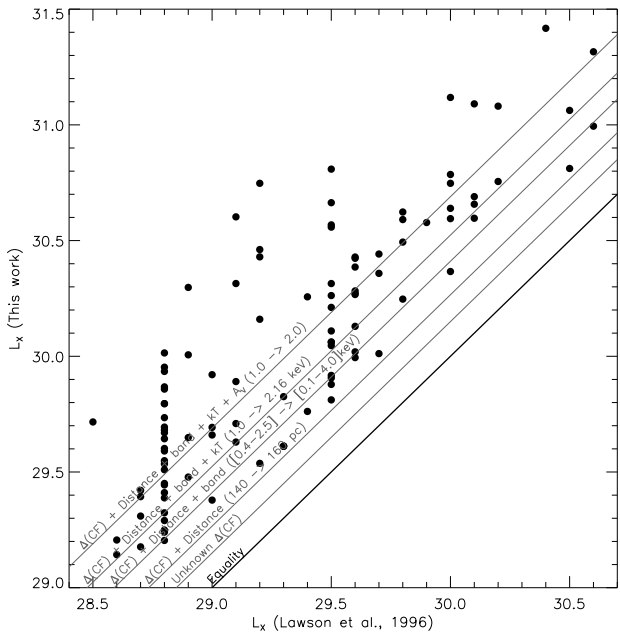}}
\caption{Comparison of X-ray luminosities reported by \citet{law96} for Chamaeleon I stars and those recomputed from the same data in this work. No distinction is made here between detections and upper limits. The bottom solid line indicates the locus of equal values; the light lines indicate the effect, on the X-ray luminosities, of: recomputing the conversion factor assuming $kT=1.0$ and $N_{\rm H}=2\cdot 10^{21}$ (see text), changing the assumed distance, source plasma temperature,  band in which $L_{\rm X}$ is computed and average extinction. The scatter of points about the highest light line is due to the adoption of individual extinction corrections.}
\label{fig:conc_cha_lxlx}
\end{figure}

In order to account for differential extinction (i.e. the fact that star are
subject to different extinctions) and to uniform our assumptions to the ONC and
NGC~2264 studies, we re-estimated X-ray luminosities, in our {\em standard}
0.1-4.0 keV band. We  started from PSPC count rates in the 0.4-2.5 keV band,
i.e. from the $L_{\rm X}$ reported in \citet{law96} divided by the above mentioned
conversion factor. We then multiplied these count-rates by conversion factors
between PSPC count-rates (in the 0.4-2.5 keV band) and luminosities (in the
0.1-4.0 keV band), computed for a $kT=2.16$ keV thermal plasma emission
absorbed by an hydrogen column $N_{\rm H}=2 \cdot 10^{21}\cdot A_{\rm V}$ and our assumed
distance to the association (160pc).  Estimates of individual optical
extinction values are taken from the following  works: \citet[][ $A_{\rm J}$, Table
3]{law96}, \citet[][ $A_{\rm V}$, Table~2]{gau92}, \citet[][ $E_{\rm B-V}$, Table
1]{wal92} and \citet[][ $A_{\rm V}$, Table~1]{cam98}; whenever multiple estimates
were available for a given star we choose one of the four values, the
precedence order being the same as the order of citation given above. $A_{\rm J}$ and
$E_{\rm B-V}$ were converted to $A_{\rm V}$ by multiplying by 3.55 and 3.1 respectively
\citep{mat90}. Figure \ref{fig:conc_cha_lxlx} compares the new X-ray
luminosities with those reported in \citet{law96} and indicates the effects
that contribute to the considerable average discrepancy between the two
estimates. First of all a difference of $\sim 0.15$ dex, indicated by the
lowest diagonal thin line, is of unclear origin: we recomputed the conversion
factor, in the 0.4-2.5keV band, assuming $kT=1.0$keV and  $N_{\rm H}=2.0\cdot
10^{21}$, i.e. following \citet{fei93}, and derived a larger conversion factor,
by $\sim 0.15$dex, respect to the value reported by these authors. The other
light lines show the effect of having changed the assumed  cluster distance, 
the chosen spectral band, the plasma temperature, and the average source
extinction. The combined effects of these changes results in our X-ray
luminosities being on average  $\sim 5$ (0.7 dex) times larger than the ones
formerly derived. 

\subsection{Activity vs. circumstellar environment}

We adopt the distinction between CTTS and WTTS presented by \citet[][ Table
B1]{law96}, excluding from our analysis 4 stars with uncertain classification,
out of our 71 with mass estimates. The distinction is based on $H_{\rm \alpha}$
emission. Our final sample comprises 28 CTTS and 39 WTTS.

\begin{figure}[t]
\centering
\resizebox{\hsize}{!}{\includegraphics{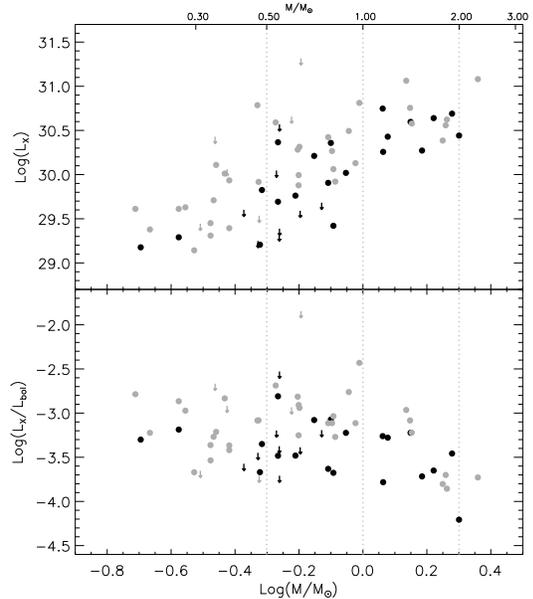}}
\caption{$L_{\rm X}$ and $L_{\rm X}/L_{\rm bol}$ vs. mass for CTTS and WTTS (black and
and gray symbols, respectively) belonging to the Chamaeleon I region. Filled circles represent detections, down-pointing arrows upper-limits.  \label{fig:act_mass_cha} }
\end{figure}

Figure \ref{fig:act_mass_cha} shows, with different symbols for CTTS and WTTS,
the scatter plots of $L_{\rm X}$ and $L_{\rm X}/L_{\rm bol}$ with mass. Disregarding for the
moment the difference between CTTS and WTTS, a trend of increasing $L_{\rm X}$ with
increasing mass, already noted by \citet{law96} and also seen in other star
forming regions,  can be clearly observed. $L_{\rm X}/L_{\rm bol}$ seems to be close to
the saturation level ($10^{-3}$) at all masses. We note that \citet{law96}, on
the basis of their lower X-ray luminosities had excluded that coronal activity
in Chamaeleon I members was saturated, contrary to what reported for other star
forming regions. Our re-analysis of the same data shows that this result can be
attributed in large part to the assumptions made in the conversion between
count-rates and X-ray luminosities and to the choice a non standard X-ray
spectral band for the calculation of $L_{\rm X}$.

\begin{figure*}[!h!]
\centering
\includegraphics[width=\textwidth]{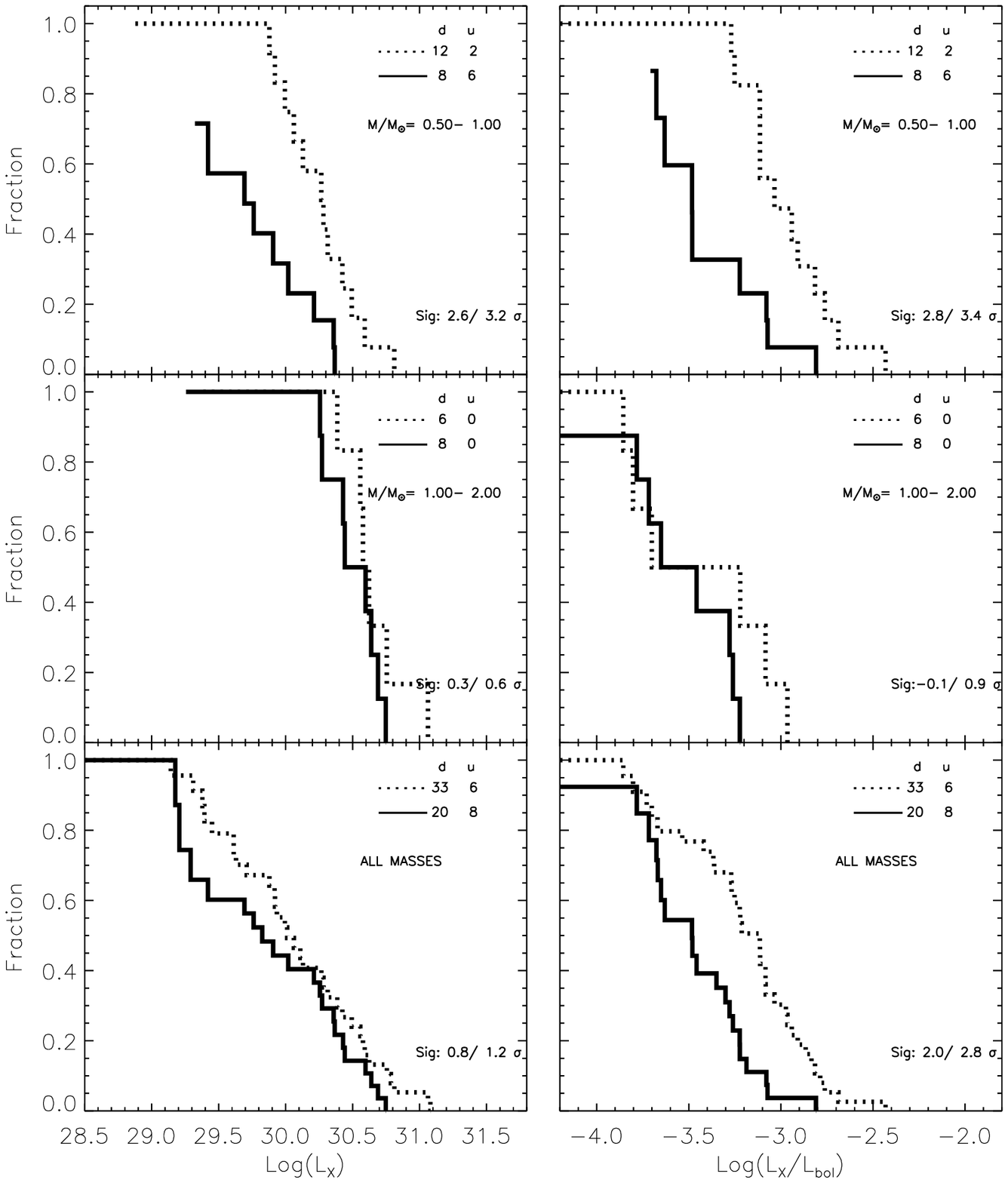}
\caption{Distributions of $L_{\rm X}$ and $L_{\rm X}/L_{\rm bol}$ (left and right columns) for CTTS and WTTS (solid
and dashed lines, respectively) in the  Chamaeleon I region. Each panel refers to a different mass range as indicated. Legends inside panels as in Fig. \ref{fig:XLF_CW_ORI}. \label{fig:XLF_CW_cha}}
\end{figure*}

Figure \ref{fig:XLF_CW_cha} shows the $L_{\rm X}$ and $L_{\rm X}/L_{\rm bol}$ distribution
functions, separately for CTTS and WTTS, in the same two mass ranges
investigated in NGC~2264 and for the whole sample. First of all we note that
there is little difference (at the $\sim 1\sigma$ level) between the two XLFs
referring to the whole population. This is indeed the same result reported by
\citet{law96}. However a look at Fig. \ref{fig:act_mass_cha} shows that this
might be due to the inclusion of stars over an ample range of masses. If we
indeed consider only stars in the $0.5-1.0M_\odot$ range CTTS appear to be
underluminous respect to WTTS at the $\sim 3\sigma$ level, both in absolute
terms and respect to their bolometric luminosities. $L_{\rm X}/L_{\rm bol}$ is indeed
lower (at the $2-2.8\sigma$ level) even if we consider the whole sample. We
obtain similar results, although of somewhat lesser significance, if we only
consider X-ray selected stars: for example, the significance of the difference
in the $0.5-1.0M_\odot$ range are $\sim 1.5$ and $\sim 2.0\sigma$ for $L_{\rm X}$ and
$L_{\rm X}/L_{\rm bol}$, respectively.

As a final note we remark that less significant results are obtained if the
same analysis is performed with the values of $L_{\rm X}$ reported by \citet{law96}.
The scatter of points around the mean relations  observed in Fig.
\ref{fig:act_mass_cha}, as well as in the distribution functions in Fig.
\ref{fig:XLF_CW_cha}, appear in this case to be larger. However the difference
in the  $0.5-1.0M_\odot$ mass range remains (at the 2.2/2.8$\sigma$ level).

\section{Summary and Conclusions \label{sect:summary}}

We have considered three SFRs: the ONC, NGC~2264 and Chamaeleon I. After a
critical re-analysis of the optical and X-ray data published in recent
literature, we have tried to answer the question of whether stars with
different circumstellar properties have different observed X-ray emission, both
in absolute terms and in relation to their bolometric luminosities. 

In all of the analyzed cases we find that CTTS are underluminous respect to
WTTS. This result is found in spite of large differences in the selection of
members and in the optical and X-ray data used. We believe that it indicates a
difference either in the intrinsic properties of X-ray emission or,
alternatively, in the radiation transport (e.g. absorption) in the proximity of
the stellar system. However we tend to prefer the first option: a difference in
the relation between optical and X-ray circumstellar extinction, for example,
might explain our result, but no such indication has been found to date.

When we could investigate the matter, i.e. in the cases of the ONC and
NGC~2264, we found that the difference holds both when we discriminate stars on
the basis of accretion and disk presence indicators. Therefore we are not able
to say which of these two related aspects is most relevant for the mechanism
responsible for the difference. 

Having established the reality of this effect, work remains to be done to
better characterize it and to identify its physical source. Thanks to sensitive
X-ray data from modern X-ray space-borne telescopes and high-throughput
optical/IR instruments this goal seems well within the reach of near future
research.

\begin{acknowledgements}
The authors wish to acknowledge support from the Italian Space Agency (ASI)
and MURST. 
\end{acknowledgements}

\end{document}